\documentclass[prb,aps,floatfix,twocolumn,superscriptaddress,footinbib,showpacs,citeautoscript]{revtex4-1}
\usepackage{graphicx}
\usepackage{epsfig}
\usepackage{braket}                                                                                                                                                                                                                                                                   
\usepackage{natbib}

\usepackage{amsfonts}
\usepackage{amsmath}
\usepackage{amssymb}
\usepackage{bm}

\begin{document}

\title{Reentrant Formation of Magnetic Polarons  in  Quantum Dots}
\author{J.~M. Pientka}
\affiliation{Department of Physics, University at Buffalo--SUNY, 
Buffalo, NY 14260}
\author{R. Oszwa\l{}dowski}
\affiliation{Department of Physics, University at Buffalo -- SUNY, 
Buffalo, NY 14260}
\author{A.~G. Petukhov}
\affiliation{Department of Physics, South Dakota School of Mines and
Technology, Rapid City, SD 57701}
\author{J. ~E. Han}
\affiliation{Department of Physics, University at Buffalo -- SUNY, 
Buffalo, NY 14260}
\author{Igor \v{Z}uti\'{c}}
\affiliation{Department of Physics, University at Buffalo -- SUNY, 
Buffalo, NY 14260}

\begin{abstract}
We propose a model of magnetic polaron formation in semiconductor quantum dots
doped with magnetic ions. A wetting layer serves as a reservoir of photo-generated 
holes, which can be trapped by the adjacent  quantum dots.
For certain hole densities, the temperature dependence of the magnetization
induced by the trapped holes is reentrant: it disappears for some temperature range
and reappears at higher temperatures. We demonstrate that this peculiar
effect is not an artifact of the mean field approximation and persists after statistical spin 
fluctuations are accounted for. We predict fingerprints of reentrant magnetic polarons 
in photoluminescence spectra.
\end{abstract}
\maketitle 

Long spin memory time~\cite{Gurung2008:APL}, giant magnetoresistance~\cite{Petukhov2000:PRB},
robust magnetic ordering, and its versatile control in epitaxial and colloidal quantum dots (QDs)%
~\cite{Beaulac2009:S,Sellers2010:PRB,Xiu2010:ACS,Fernandez-Rossier2004:PRL,
Govorov2005:PRB,Gall2011:PRL,Abolfath2008:PRL,Bussian2009:NM} 
are attributed to magnetic polarons (MPs), known for fifty years in bulk 
semiconductors~\cite{Yakovlev:2010}. The MP formation
 can be viewed as  a ``cloud'' 
of localized spins, aligned through exchange interaction with a confined carrier spin. 
While the seminal studies of MPs in the  bulk~\cite{Dietl1983:PRB,Wolff:1988, Nagaev:1983} 
assumed that an impurity binds only one carrier, many experiments
in QDs demonstrate multiple occupancies%
~\cite{Cherenko2010:PSSB,Matsuda2007:APL,Bansal2009:PRB,Fisher2005:PRL,
Besombes2005:PRB}.
We show that varying QD occupancy has important consequences for MP formation.  
It is conventionally understood that MPs form at low temperature ($T$)  and vanish at high $T$, 
owing to thermal fluctuations of Mn spins. Here, we propose an unexpected 
scenario where the temperature can {\em enhance}, rather than quench, 
MP formation and  lead to reentrant magnetism~\cite{Petukhov2007:PRL,reentrant}.

We formulate a model of MPs formed in II-VI semiconductor QDs doped with Mn ions 
and apply it to simulate photoluminescence (PL) 
spectra~\cite{Maksimov2000:PRB,Seufert2002:PRL,%
Sellers2010:PRB,Beaulac2009:S, Takeyama1995:PRB}. In a PL 
experiment~\cite{Zutic2004:RMP}, the number of carriers captured by QDs depends
on the total density of the photoexcited carriers, determined by the laser intensity. 
Focusing on holes in type-II QDs~\cite{Sellers2010:PRB},   
we find nontrivial dependencies of MP binding energies on both the laser intensity and $T$.
\begin{figure}[tbph]
\centerline{\psfig{file=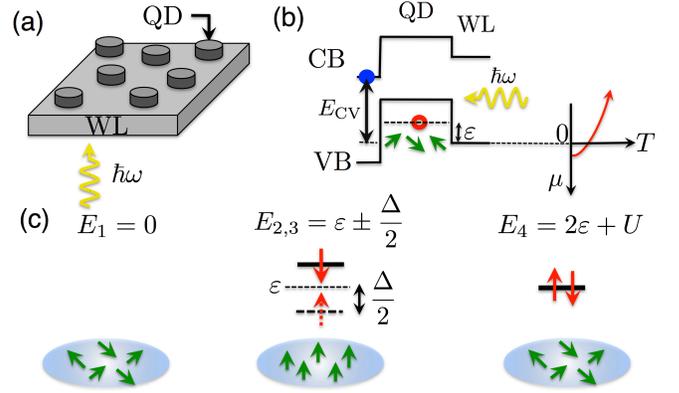,width=1\linewidth,angle=0}}
\caption{(Color online)~(a) A scheme of QDs grown on a wetting layer (WL), which is
excited by light (cladding layer not shown).  Type-II conduction/valence band (CB/VB) 
profile of a II-VI QD doped with Mn spins (green).  
$\varepsilon$ and $E_{CV}$ are the confinement and band gap energies. 
Hole quasi-Fermi level $\mu(T)$ lies in the continuum of WL states for $T=0$. 
(c) QD states with corresponding energies,
occupancies, and Mn-spin alignment:  $E_1$ state has zero energy and QD occupancy;
$E_{2,3}$ state has one hole with spin $+3/2$ ($-3/2$), $\Delta$ is the
exchange splitting; and $E_4$ has two holes with a repulsive Coulomb energy $U$.
 }
\label{fig:Fig1}
\end{figure}

Figure~\ref{fig:Fig1} shows our model.  
Epitaxial QDs typically reside on a two-dimensional (2D) wetting layer (WL) in which 
electron-hole pairs are created by interband absorption of light~\cite{NoteHoles}.
Under illumination, 
a quasi-Fermi distribution is established in the WL 2D hole gas.  
Since the total area of the QDs is low relative 
to that of the WL,  the quasi-Fermi level, 
$\mu(T,p)= + k_{B}T\ln \left(\exp\left[ p \pi\hbar^2/(m^*_h k_BT )\right] -1 \right) $, 
is pinned by the non-equilibrium WL holes with 2D density $p$, where $m^*_h$ is the heavy-hole
effective mass. The population of the QDs by the captured holes is  controlled by $\mu$.
This assumption is justified by the fact that in many QD systems, the times of capture and intra-dot
relaxation are much shorter than carrier radiative recombination time. Therefore, quasi-equilibrium
can be established in the valence band~\cite{NoteQuassi}.

The exchange interaction of heavy holes
with Mn spins in a flat QD is highly anisotropic and
described by an Ising Hamiltonian~\cite{Gall2011:PRL},  
\begin{equation}
H_{\text{ex}}= -\beta\sum_{i,j}\delta({\bf r}_i -{\bf R}_j) s_{zi} S_{zj} ,\label{eq:Hex}
\end{equation}
where $\beta$ is the exchange coupling constant,  ${\bf r}_i$ $({\bf R}_j)$ and $s_{zi}$ $(S_{zj})$ 
are the position and the spin projection of the 
carrier (Mn)~\cite{Furdyna1988:JAP}.
The exchange interaction results in the Mn-magnetization, $M_z$, and exchange splitting $\Delta$ of hole levels. 
We assume uniform hole wavefunction throughout the 
QD volume $\Omega$~\cite{NoteWave},
thus we can relate $M_z$ and $\Delta= \beta M_z/ g \mu_{B}$.  
 The maximum $M_z$ is  $M_{\text{max}}= x_{\text{Mn}}N_0\Omega S g \mu_B$,  
where 
$x_{\rm Mn}$ is the Mn fraction per cation,
$N_0$ is the density of cation sites, $S = 5/2$, $g$ = 2.0 is the 
$g$ factor,  $\mu_B$ is the Bohr magneton,
and $N_0 \beta \sim -1.0 \text{ eV} $~\cite{Furdyna1988:JAP}.

The Gibbs free energy of the system is expressed  as 
\begin{equation}
G_{\text{sys}}(\xi) = G_{\text{Mn}}(\xi)+F_{h}(\xi),
 \label{eq:Fsys}%
\end{equation}
in terms of the order parameter is  $\xi=M_z/M_{\text{max}}$,  where 
$G_{\text{Mn}}(\xi)$ is the Mn-spins contribution, 
and $F_h(\xi)$ is the hole grand canonical (GC)  free energy.
$G_{\text{Mn}}(\xi)$ can be obtained by expressing the free energy of the Mn spins 
as a function of an external magnetic field and then applying a 
Legendre transformation:~\cite{Petukhov2007:PRL,Kubo:1960}
\begin{eqnarray}
\!\!\!\!G_{\text{Mn}}(\xi)&=&k_{B} T N_{\text{Mn}}  \nonumber \\
&\times & \left [ \xi B_S^{-1}(\xi)-
\ln\frac{\sinh\left[ (1+1/2S) B_S^{-1}(\xi) \right]  }{\sinh\left[ B_S^{-1}(\xi)/2S
 \right]} \right],
\label{eq:MnFree}
\end{eqnarray}
where $B_S^{-1}(\xi)$ is  the inverse Brillouin function~\cite{Kubo:1960}, 
and $N_{\rm Mn}$ the number of Mn in the QD.  
$F_{h} (\xi) $ is obtained from the QD states with 0, 1 or 2 holes [see Fig.~\ref{fig:Fig1}(c)].  
For transparency, we consider only one nonmagnetic single-hole level and
neglect the possibility of magnetization in the presence of 
two holes~\cite{Fernandez-Rossier2004:PRL,NoteMBP}.
This yields 
\begin{eqnarray}
F_{h}(\xi)&=& -k_{B} T \ln \left[   1 + 2e^{- \left(  \varepsilon-\mu\right)  
/k_{B} T }\cosh\left[\frac{\Delta_{\text{max}} \xi}{2 k_B T} \right] \right.  \nonumber \\
&+& \left. e^{- \left(  2\varepsilon+
U- 2 \mu\right)  /k_{B} T} \right],\label{eq:hFree}%
\end{eqnarray}
where $\varepsilon$ is the single-hole confining energy, $U$ is the repulsive Coulomb (charging) energy, 
and the maximum splitting is $\Delta_{\text{max}}= x_{\text{Mn}}|N_0\beta| S$. 
To elucidate some interesting phenomena, we first use a standard mean field (MF) 
approximation~\cite{Wolff:1988,Nagaev:1983,Dietl1983:PRB}, commonly used by many authors~\cite{Fernandez-Rossier2004:PRL,Govorov2005:PRB,Wojnar2007:PRB,Beaulac2009:S}. 
By numerically minimizing Eq.~(\ref{eq:Fsys}), we obtain the value $\xi_{\text{MF}}$.
The MP energy, $E_{\text {MP}}$, 
is defined as the expectation value of Eq.~(\ref{eq:Hex}).  
We calculate it from $E_{\text{MP}} = -k_B T 
\Delta_{\text{max}} d \left( \ln\mathcal{Z}_{\text{sys}}\right)/d\Delta_{\text{max}}$~\cite{Bhattacharjee1986:SSC}. 
The MF partition function 
$\mathcal{Z}_{\text{sys}}\left( \xi_{\text{MF}}\right)=
\mathcal{Z}_{\text{Mn}}\left( \xi_{\text{MF}}\right) \mathcal{Z}_{h}\left( \xi_{\text{MF}}\right)$ 
is expressed in terms of the Mn and hole contributions:
$\mathcal{Z}_{\text{Mn}}\left( \xi_{\text{MF}}\right)=e^{-G_{\text{Mn}}(\xi_{\text{MF}})/k_B T}$,
$\mathcal{Z}_{h}\left( \xi_{\text{MF}}\right)= e^{-F_h(\xi_{\text{MF}})/k_B T}$.  
We obtain
\begin{equation}E_{\text{MP}}= -\frac{ \Delta_{\text{max}}\xi_{\text{MF}}}{\mathcal{Z}_h(\xi_{\text{MF}})}
e^{- \left(  \varepsilon-\mu\right)  
/k_{B} T }\sinh \left[\frac{\Delta_{\text{max}}\xi_{\text{MF}}}{2k_{B} T}\right].
\label{eq:EmpMF}%
\end{equation}

\begin{figure}[t]
\centerline{\psfig{file=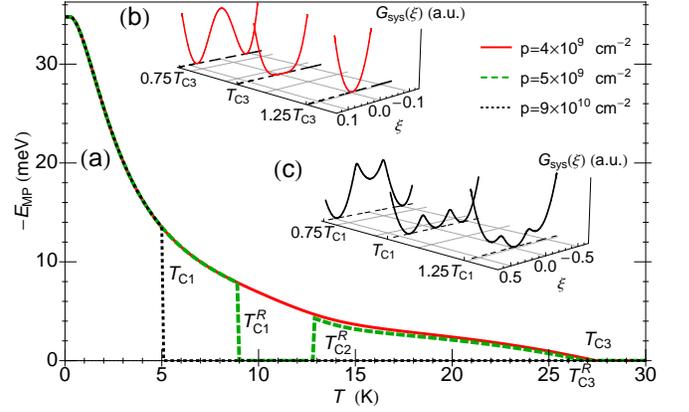,width=1.0\linewidth,angle=0}} 
\caption{(Color online)~(a)  Reentrant behavior in the MP energy, $see E_{\text{MP}}$ [Eq.~(\ref{eq:EmpMF})],
and various critical temperatures.
Free energy evolution reveals the second-  
(b) and first-order (c) phase transitions
for $p=4\times10^9 $ and  $p=9\times10^{10}\text{ cm}^{-2}$, respectively. 	 
Parameters, partially guided by (Zn,Mn)Te/ZnSe QDs~\cite{Sellers2010:PRB}:
$\varepsilon$= $-$36 meV, $U=30$ meV~\cite{Matsuda2007:APL,Note2}, 
$\Delta_{\text{max}}=69 \text{ meV}$ (see Fig.~\ref{fig:Fig1}), 
$x_{\text {Mn}}$=2.6\%, 
$N_0\beta$ = $-$1.05 eV, and $m_h^* = 0.21$.  QD volume, $\Omega$, 
is cylindrical with a 50~\AA ~radius and a 25~\AA ~height. }
\label{fig:Fig2}
\end{figure}

In Fig.~\ref{fig:Fig2}, MF predicts multiple phase transitions and reentrant magnetism. 
For low-hole densities (solid line), the system exhibits a second-order phase transition at  $T_{C3}=27$ K. 
In Fig.~\ref{fig:Fig2}(b), the location of minimum $G_{\text{sys}}(\xi)$ continuously goes to zero, a signature 
of  second order transition. This is similar to the usual MP case (single particle), since for any $T<30$ K, the probability 
of finding a single hole in the QD is dominant.
For high-hole densities (dotted line), 
MF predicts a first-order transition at $T_{C1}=5$ K, consistent with the discontinuous shift of  the 
 $G_{\text{sys}}(\xi)$ minimum  to $\xi=0$ at $T_{C1}$ in Fig.~\ref{fig:Fig2}(c). 
At $T<T_{C1}$, magnetism is present since the QD is occupied by 1,
rather than 2 holes (despite  $\mu \sim 1 \text{ meV}$ in the continuum), according to 
$E_2 - \mu < E_4 - 2 \mu $  [Fig.~\ref{fig:Fig1}(c)]. 
This inequality is satisfied as long as the ordering of Mn-spins sufficiently lowers the energy
of the single hole state $E_2$.
 When $T>T_{C1}$, $E_2- \mu > E_4 - 2 \mu $ and  the QD becomes doubly occupied suppressing MP formation. 
Above $T_{C1}$, magnetism does not reappear 
since thermal Mn excitations completely quench magnetic order, before $\mu$ approaches $\varepsilon +U$ to promote single occupancy. 

The QD exhibits reentrant magnetism at $p = 5 \times10^{9} \mbox{ cm}^{-2}$ (dashed line).
For $T<T^R_{C2}$, the scenario is the same as the dotted line ($T^R_{C1}$ plays the role of $T_{C1}$).  
At $T> T^R_{C2}$, with $E_2 - \mu < E_4 - 2 \mu $, the QD is singly occupied as a result of  $\mu$ moving quickly towards $\varepsilon$, 
thus MP reappears by a first-order transition. The transition at $ T=T_{C3}^R$ has the same origin as the solid curve.  For $ T\gg T_{C3}^R$, 
the dot becomes emptied. 

However,  MF theory neglects the possibility for the system to deviate from the equilibrium value, 
$\xi_{\text{MF}}$.  This leads to unphysical thermodynamic phase transitions 
in small systems.  Could this also imply that the described reentrant magnetism is only
an artifact of the MF theory? To address this question and better understand 
the validity of the behavior predicted at the MF level, we formulate a fluctuation 
approach (FA).  Statistical fluctuations are included in the partition function by integrating 
over all possible values of the order parameter~\cite{Dietl1983:PRB}.  Correspondingly, 
we employ
 $\mathcal{Z}_{\text{sys}} =\int_{-1}^{1}  e^{-G_{\text{sys}}(\xi)/k_B T} d\xi$
to implement the framework used for MF [recall 
 Eq.~(\ref{eq:EmpMF})], and obtain
the average exchange energy
\begin{eqnarray}
E_{\text{MP}} &=&-\frac{ \Delta_{\text{max}}}{\mathcal{Z}_{\text{sys}}}
e^{- \left(  \varepsilon-\mu\right)  /k_{B} T }   \nonumber \\
&\times&\int_{-1}^{1} \!\! d\xi \ \xi e^{- G_{\text{Mn}}\left(\xi \right)  /k_{B} T }   
\sinh \left[\frac{\Delta_{\text{max}}\xi}{2k_{B} T}\right].
\label{eq:EMPFM}
\end{eqnarray}
This $E_{\text{MP}}$ for the GC~ensemble is similar to that 
of the canonical ensemble~\cite{Umehara2000:PRB}.  However, $\mathcal{Z}_{\text{sys}}$ 
now contains multiple occupancies  and the numerator is weighted by $e^{\mu /k_BT}$, 
which decreases with increasing 
 $T$ [Fig.~\ref{fig:Fig1}(b)].  
 
\begin{figure}[tbph]
\centerline{\psfig{file=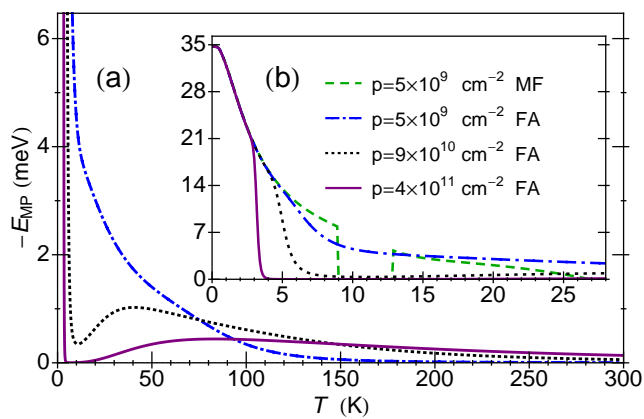,width=1.0\linewidth,angle=0}} 
\caption{ (Color online)~(a) Reentrant magnetism in the fluctuation approach (FA) [see Eq.~(\ref{eq:EMPFM})] 
for $p= 5 \times  10^9,$ $9\times10^{10},$  $4\times10^{11} \text{ cm}^{-2}$. 
(b) Comparison of the FA and mean field (MF).  FA (dot-dashed line)  shows no 
reentrant magnetism predicted for MF and  $p=5\times10^{9} \text{ cm}^{-2}$ (dashed line).
Other QD parameters are from Fig.~\ref{fig:Fig2}.} 
\label{fig:Fig3}
\end{figure}

We are now able to directly compare MF and FA results.  
The sharp MF  phase transitions [$E_{\text{MP}}$ in Fig.~\ref{fig:Fig2}], 
become smeared out, as seen in Fig.~\ref{fig:Fig3}.
FA yields finite  $E_{\text{MP}}$ at any finite $T$. 
This is expected from averaging of $G_{\text{sys}}$ over  $\xi$, implicit in Eq.~(\ref{eq:EMPFM}), 
including strong contributions from the competing local minima [Fig.~\ref{fig:Fig2}(c) 
at $T_{C1}$]. For example, 
MF reentrant magnetism from Figs.~\ref{fig:Fig2}(a) and~\ref{fig:Fig3}(b) for 
$p=5\times10^{9} \text{ cm}^{-2}$ is absent in FA (see Fig.~\ref{fig:Fig3}), since the local minima of $G_{\text{sys}}$
at $|\xi| > 0$ contribute strongly in the temperature range of 9$-$13 K. Surprisingly,  
the dotted and solid curves in Fig.~\ref{fig:Fig3} show FA reentrant $E_{\text{MP}}$ even at room temperature, 
while for the same $p$ no MF reentrant behavior was seen. In FA, the increase in 
$|E_{\text{MP}}|$ at higher $T$ is due to the the QD occupancy
decreasing from 2 to 1 holes.  
The maximum reentrant $|E_{\text{MP}}(T)|$ coincides with the average occupancy of  1.  
Even though inclusion of statistical fluctuations removes reentrance for some hole 
densities, it also yields a smoothed version of the same effect for higher $T$, where MF predicts 
$E_{\text{MP}}=0$.  A peculiar nonmonotonic $E_{\text{MP}}(T)$ is therefore not limited to the MF description.  

Furthermore, the reentrant MP is not restricted to the above parameters, but occurs for a range of  
$x_{\text{Mn}}$, $\varepsilon$, $U$, and $p$.  For example, we find reentrant $E_{\text{MP}}$  
for $1.5 \times 10^{10}\text{ cm}^{-2} \leq p \leq 2 \times 10^{12}\text{ cm}^{-2}$, with other 
parameters fixed. Conversely, the reentrant MP is present for 
$1.5 \% \leq x_{\text{Mn}} \leq 2.6 \%$,  if the remaining parameters are fixed.

We next discuss how reentrant  magnetism could be observed in PL experiments. To obtain the 
PL spectrum, we assume low QD density and sufficiently high intensity of exciting light for 
two CB electrons to always be in the vicinity of each QD, occupying the lowest possible 
energy $\sim E_{\text{CV}} >0$~\cite{Bacher2002:PRL} (see Fig.~\ref{fig:Fig1}) and with opposite spins.
The total spectrum, $I_{\text{tot}}$ is the superposition of the lines generated by the $2 \rightarrow 1$,
 and $1 \rightarrow 0$ transitions,
 \begin{equation}
I_{\text{tot}} (\omega) = I_{1 \rightarrow 0} (\omega) + I_{2 \rightarrow 1} (\omega).
\label{eq:Isys}%
\end{equation}

We assume that the Mn-configuration does not change during a recombination event.  The intensity of 
each PL line is 
$I  =\sum_{i,f} p_i w_{if} \delta\left[\hbar \omega -( E_i -E_f)\right]$, where $w_{if}$ 
is the transition rate, $p_i$ is the thermodynamic probability of the initial state, $\hbar \omega$ 
is the energy of the emitted photon, and $E_f$ ($E_i$) is the energy of the final (initial) state of the system.
We replace the above $\sum$ with $\int  d\xi$.
For 1 $\rightarrow$ 0 transitions, the system is in an initial state with a hole of spin up (down), 
which later recombines with a spin-down (up) electron~\cite{Wojnar2007:PRB}.  
The intensity of this line is
\begin{eqnarray}
I_{1 \rightarrow 0} (X)&=& c(T)e^{- \left(  \varepsilon-\mu\right)  /k_{B} T } e^{- X /k_{B} T } \nonumber \\
&\times &e^{-G_{\text{Mn}}(2X/\Delta_{\text{max}})/k_B T} \theta \left( \Delta_{\text{max}} - |2 X| \right).
\label{eq:PL10}%
\end{eqnarray}
Here, $X(\omega) =  \hbar \omega -E_{\text{CV}} -\varepsilon$ is the shifted frequency,
 $c(T)\propto (\Omega/\beta) |d_{cv}|^2/\mathcal{Z}_{\text{sys}}$,  $d_{cv}$ the dipole matrix element,
 and $\theta$ is the step function.
For the $2\rightarrow1$ transition, there are two holes and two electrons of opposite spin in the initial state.  
The intensity is 
\begin{eqnarray}
I_{2 \rightarrow 1} (X)&=& c(T) e^{- \left(  2\varepsilon+U-2\mu\right)  /k_{B} T }
e^{-G_{\text{Mn}}\left( 2(X-U)/\Delta_{\text{max}} \right)/k_B T} \nonumber \\
&\times& \theta \left( \Delta_{\text{max}} - 2| X-U| \right).
\label{eq:PL21b}%
\end{eqnarray}

\begin{figure}[tr]
\centerline{\psfig{file=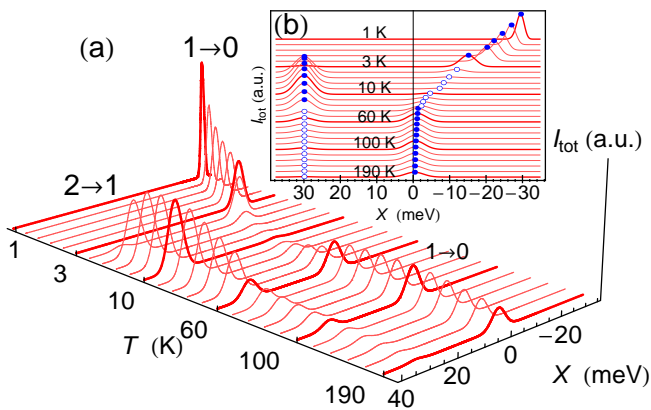,width=1.\linewidth,angle=0}}
\caption{(Color online)~(a) A PL spectrum for reentrant magnetism showing $2 \to 1$ 
and $1\to 0$ hole occupancy 
transitions ($p=9\times10^{10}$ cm$^{-2}$). The thick lines emphasize important features in the PL 
spectrum, while the thin lines between them show their evolution at intermediate equidistant $T$. 
(b) Overall PL peak position marked with filled circles.
As $T$ increases, $1 \to 0$ peak shifts to zero energy. QD parameters are 
from Fig.~\ref{fig:Fig2}.}
\label{fig:Fig4}
\end{figure}

In Fig.~\ref{fig:Fig4}, the PL spectrum shows the evolution of the peaks
for transitions $1 \to 0$, centered at $X<0$, and $2 \to 1$,
centered at the charging energy, $X=U$, since 
$\hbar \omega=E_i-E_f = (2 \varepsilon + U + E_{\text{CV}})-\varepsilon$ ~\cite{Note4}.
From Eqs.~(\ref{eq:MnFree}), (\ref{eq:PL10}), and  (\ref{eq:PL21b}) it follows 
that the Mn-contribution to the PL 
is $T$ independent: the amplitude of the
$1 \to 0$ ($2 \to1$)
 peak at different $T$ 
is proportional to the probability of finding a single (double) occupied QD.
For $T<5$ K, the $1\rightarrow0 $ line dominates, while it becomes 
negligible at $\sim 10$ K where the system is virtually nonmagnetic
and  the double occupied state ($2\rightarrow1$ line) is dominant. At higher $T$,  
due to the shift of $\mu$ toward CB, the probability of single occupancy increases,
and for $T> 100$ K, the probability of zero occupancy increases. 
The resulting $T$-dependencies are remarkably nonmonotonic 
for both $I_{\text{tot}}$ peak position (red and blue  shifts), 
and $I_{\text{tot}}$ peak intensity; a signature of reentrant MPs, 
consistent with $E_\text{MP}(T)$ in Fig.~\ref{fig:Fig3}.

What are the semiconductor systems where  
the reentrant MP could be found?~Recently, a non\-mono\-tonic PL red shift 
was observed in type-II  (Zn,Mn)Te/ZnSe QDs~\cite{Sellers2010:PRB}, 
which have partially guided our choice of  parameters. 
However, the reentrant magnetism should not be limited to type-II systems.
The necessary condition is the $T$-dependent multiple occupancy, 
readily seen in both type-II~\cite{Matsuda2007:APL,Bansal2009:PRB}
and type-I QDs~\cite{Cherenko2010:PSSB,NoteType1}.
Multiple occupancy can be reached by raising photo-excitation intensity, which
may first lead to weakening of MPs (blue shift) through Mn-spin heating%
~\cite{Maksimov2000:PRB,Schoemig2003:JS,Bacher2005:PhysE,Worschech2008:SST}.
Nevertheless, an \emph{increased} blue shift was attributed to double 
occupancy in type-I magnetic QDs~\cite{Schoemig2003:JS}.  

Considering only a reduced $T$-range could conceal the presence of reentrant MP.
An initial steep decline in $E_{\text{MP}}(T)$ (see Fig.~\ref{fig:Fig3}) is similar to conventional MPs,
while a slightly higher $T$ region (e.g., 5 to 30 K for solid line in Fig.~\ref{fig:Fig3}) 
could be misinterpreted as a final thermal breakup of MPs.  
Thus, further experimental studies of power and $T$ dependence are important. 
Specifically, it would be desirable to consider single self-assembled 
QDs~\cite{Gall2011:PRL, Klopotowski2011:PRB} to reduce uncertainties due to 
inhomogeneous averaging, and focus on moderate $x_{\rm {Mn}}$ 
to suppress the Mn-Mn antiferromagnetic interactions~\cite{Sellers2010:PRB}.
Colloidal (II,Mn)VI QDs~\cite{Beaulac2009:S,Bussian2009:NM} 
showing a robust MP formation are also promising candidates 
to test some of our predictions.

Even though in this work we have only focused on reentrant magnetism, we expect that further studies of the implications of multiple occupancies will lead to additional surprises in both epitaxial and colloidal QDs. Since prior findings in QDs were successfully applied to different finite fermion systems  \cite{Reimann2002:RMP}, it may be possible to seek other promising paths for observing reentrant magnetism.

We thank A. Petrou and K. Vyborny for valuable discussions.
This work was supported by DOE-BES, AFOSR-DCT, U.S. ONR, NSF-NRI NEB 2020, SRC, and NSF-DMR.


\begin{thebibliography}{10}
\bibitem{Gurung2008:APL}
T. Gurung, S. Mackowski, G. Karczewski, H.~E. Jackson, and L.~M. Smith,
Appl. Phys. Lett. {\bf 93}, 153114 (2008).

\bibitem{Petukhov2000:PRB}
 A.~G. Petukhov and M. Foygel, Phys. Rev. B {\bf 62}, 520 (2000).

\bibitem{Beaulac2009:S}
R.  Beaulac, L. Schneider, P.~I. Archer, G. Bacher, and D.~R. Gamelin,
Science {\bf 325}, 973  (2009).

\bibitem{Sellers2010:PRB}
I.~R. Sellers, R.~Oszwa\l{}dowski, V.~R. Whiteside, M.~Eginligil, A.~Petrou, I.~\v{Z}uti\'{c}, 
W.-C. Chou, W.~C. Fan, A.~G. Petukhov, S.~J. Kim, A.~N. Cartwright, and B.~D. McCombe,
Phys. Rev. B  {\bf 82}, 195320 (2010).

\bibitem{Xiu2010:ACS}
F. Xiu, Y. Wang, J. Kim, P. Upadhyaya, Yi Zhou, X. Kou, W. Han, 
R.~K. Kawakami, J. Zou, and K.~L. Wang.
ACS Nano  {\bf 4}, 4948  (2010), and references therein.

\bibitem{Fernandez-Rossier2004:PRL} 
J.~Fern\'andez-Rossier and L.~Brey,
Phys. Rev. Lett. {\bf 93}, 117201 (2004).

\bibitem{Govorov2005:PRB}
A.~O. Govorov,
Phys. Rev. B.  {\bf 72}, 075359 (2005);  W. Zhang, T. Dong, and A.~O. Govorov,
Phys. Rev. B.  {\bf 76}, 075319 (2007).

\bibitem{Gall2011:PRL}
C. Le Gall, A. Brunetti, H. Boukari, and L. Besombes
Phys. Rev. Lett. {\bf 107}, 057401 (2011).

\bibitem{Abolfath2008:PRL}
R.~M. Abolfath, P. Hawrylak, and I. \v{Z}uti\'c, 
Phys. Rev. Lett. {\bf 98}, 207203 (2007);
R.~M. Abolfath, A.~G. Petukhov, and I. \v{Z}uti\'c, 
Phys. Rev. Lett. {\bf 101}, 207202 (2008);
N.~T.~T. Nguyen and F. M. Peeters,
Phys. Rev. B {\bf 78}, 045321 (2008);
Y. G. Semenov and K.~W. Kim ,
Phys. Rev. B {\bf 72}, 195303 (2005);
N. Lebedeva, A. Varpula, S. Novikov, and P. Kuivalainen,
Phys. Rev. B {\bf 81}, 235307 (2010);
L. Villegas-Lelovsky, F. Qu, L.~O. Massa, V. Lopez-Richard,  and G.~E. Marques
Phys. Rev. B {\bf 84}, 075319 (2011);
S.-J. Cheng,
Phys. Rev. B {\bf 77}, 115310 (2008). 

\bibitem{Bussian2009:NM}
D.~A. Bussian, S.~A. Crooker, M. Yin, M. Brynda, A.~L. Efros, and V.~I. Klimov, 
Nature Mater. {\bf 8}, 35 (2009).

\bibitem{Yakovlev:2010}
D.~R. Yakovlev and W.Ossau,  in
{\em Introduction to the Physics of Diluted Magnetic Semiconductors}
edited by J. Kossut and J.~A. Gaj (Springer, Berlin, 2010).

\bibitem{Dietl1983:PRB}
T.~Dietl and J.~Spa\l{}ek, Phys. Rev. B  {\bf 28}, 1548  (1983).

\bibitem{Wolff:1988} 
P.~A. Wolff,  in  {\em Semiconductors and Semimetals} edited by J.~K. Furdyna and
J. Kossut  (Academic Press, San Diego 1988), Vol. 25.

\bibitem{Nagaev:1983}   
E.~L. Nagaev,  {\em Physics of Magnetic Semiconductors}
(MIR Publishers, Moscow 1983).

\bibitem{Cherenko2010:PSSB}
A.~V. Chernenko, A.~S. Brichkin, S.~V. Sokolov, and S.~V. Ivanov,
Phys. Status Solidi B {\bf 247}, 1514 (2010).

\bibitem{Matsuda2007:APL} 
K.~Matsuda, S.~V. Nair, H.~E. Ruda, Y.~Sugimoto, T.~Saiki, and K.~Yamaguchi,
Appl. Phys. Lett. {\bf 90}, 013101 (2007).

\bibitem{Bansal2009:PRB}
B. Bansal, S.~Godefroo, M.~Hayne, G.~Medeiros-Ribeiro, and V.~V. Moshchalkov,
Phys. Rev. B {\bf 80}, 205317 (2009).

\bibitem{Fisher2005:PRL}
B. Fisher, J.~M. Caruge, D. Zehnder, and M.~Bawendi, 
Phys. Rev. Lett.  {\bf 94}, 087403 (2005).

\bibitem{Besombes2005:PRB}
L.~Besombes, Y.~Leger, L.~Maingault, D.~Ferrand, H.~ Mariette, and 
J.~Cibert,
Phys. Rev. B {\bf 71}, 161307(R) (2005).
 
\bibitem{Petukhov2007:PRL}
A.~G. Petukhov, I. \v{Z}uti\'c, and S.~C. Erwin,
Phys. Rev. Lett. {\bf 99}, 257202 (2007).

\bibitem{reentrant}
In the bulk, reentrant magnetism could be possible due to large thermally-excited 
carrier density [Ref.~\onlinecite{Petukhov2007:PRL}, 
M.~J. Calderon and S. {Das Sarma}, Phys. Rev. B {\bf 75}, 235203 (2007), and experiments
of V. Krivoruchko, V. Tarenkov, D. Varyukhin, A. DÕyachenko, O. Pashkova, and V. Ivanov, 
J. Magn.  Magn. Mater. {\bf 322}, 915 (2010)] 
or spin glass formation [C.~K. Thomas, D.~A. Huse, and A. ~A. Middleton,
Phys. Rev. Lett. {\bf 107}, 047203 (2011)].

\bibitem{Maksimov2000:PRB}
A.~A. Maksimov, G.~Bacher, A.~McDonald, V.~D. Kulakovskii, A.~Forchel, C.~R. Becker, 
G.~Landwehr, and L.~W. Molenkamp,
Phys. Rev. B {\bf 62}, R7767 (2000).

\bibitem{Seufert2002:PRL}
J.~Seufert, G.~Bacher, M.~Scheibner, A.~Forchel, S.~Lee, M.~Dobrowolska, and J.~K. Furdyna,
Phys. Rev. Lett. {\bf 88}, 027402 (2001).

\bibitem{Takeyama1995:PRB} 
S.~Takeyama, S.~Adachi, Y.~Takagi, V.~F. Aguekian,
Phys. Rev. B {\bf 51}, 4858 (1995).

\bibitem{Zutic2004:RMP}
I. \v{Z}uti\'{c}, J. Fabian, and S.~Das~Sarma,
Rev. Mod. Phys. {\bf 76}, 323 (2004); J. Fabian,  A. {Matos-Abiague}, 
C. Ertler, P. Stano, and I. \v{Z}uti\'c, Acta Phys. Slov. {\bf 57}, 565 (2007);
S. Maekawa {\em Concepts in Spin Electronics} (Oxford University Press, Oxford, 2006).
\bibitem{NoteHoles}
For simplicity,  we consider here only  heavy holes.

\bibitem{NoteQuassi}
R.~Heitz, M.~Veit, N.~N. Ledentsov, A.~Hoffmann, D.~Bimberg, V.~M. Ustinov, P.~S. Kop'ev, and Zh.~I. Alferov,
Phys. Rev. B  {\bf 56}, 10435 (1997);  
I.~V. Ignatiev, I.~E. Kozin, S.~V. Nair, H.~W. Ren, S.~Sugou, Y.~Masumoto,
Phys. Rev. B  {\bf 61}, 15633 (2000); 
M.~Funato, K.~Omae, Y.~ Kawakami, Sg.~Fujita, C.~Bradford, A.~ Balocchi, A.~Prior, and B.~C. Cavenett,
Phys. Rev. B  {\bf 73}, 245308 (2006);
S. Mackowski, G. Prechtl, W. Heiss, F. Kyrychenko, G. Karczewski, J. Kossut
Phys. Rev. B  {\bf 69}, 205325 (2004);
The recombination time is even longer in type-II QDs \cite{Sellers2010:PRB}.
F. ~Hatami, M.~Grundmann, N.~N. Ledentsov, F. ~Heinrichsdorff, R.~ Heitz, J.~B\"ohrer, D.~Bimberg ,
S.~S. Ruvimov, P.~Werner, V.~M. Ustinov, P. S.~Kop'ev, and Zh.~I. Alferov.
 Phys. Rev. B  {\bf 57}, 4635 (1998).
\bibitem{Furdyna1988:JAP}
J.~K. Furdyna, J. Appl. Phys.  {\bf 64} R29 (1988). 
\bibitem{NoteWave}
A. Golnik, J.~Ginter, and J.~A. Gaj,
J. Phys. C{\bf 16}, 6073 (1983).
A uniform wave function is a good approximation.  When carrier 
localization effects are considered, using a different wave function and the canonical ensemble,
we find at most a 15 \% change in $E_{\text{MP}}$
at $T$= 5~K.

\bibitem{Kubo:1960}
R.~Kubo, {\em Statistical Mechanics} (North-Holland, Amsterdam, 1960).


\bibitem{NoteMBP}
We have determined that the effects of magnetic bipolarons, described in R. Oszwa\l{}dowski, I. \v{Z}uti\'{c}, and A.~G. Petukhov,
Phys. Rev. Lett.  {\bf 106}, 177201 (2011),
are only important at very low $T$, much below the non-magnetic doubly occupied regime.
All our main findings here are thus not altered. 

\bibitem{Wojnar2007:PRB}
P.~Wojnar, J.~Suffczy\ifmmode~\acute{n}\else \'{n}\fi{}ski, K.~Kowalik,
A.~Golnik, G.~Karczewski, and J.~Kossut.
Phys. Rev. B.  {\bf 75}, 155301 (2007).

\bibitem{Note2}
A much larger $U$ would allow only one carrier to be confined, and would
require a simpler form of $F_h$ in Eq.~(\ref{eq:hFree}). 

\bibitem{Bhattacharjee1986:SSC}
A. Bhattacharjee,  
Solid State Commun. {\bf 57}, 31 (1986).

\bibitem{Umehara2000:PRB}
M. Umehara,
Phys. Rev. B. {\bf 61}, 12209 (2000).

\bibitem{Bacher2002:PRL}
This quantity may be $T$-dependent owing to the Varshni shift; 
G. Bacher, A.~A. Maksimov, H. Sch{\"o}mig, V.~D. Kulakovskii, M.~K. Welsch,
A. Forchel, P.~S. Dorozhkin, A.~V. Chernenko, S. Lee, M. Dobrowolska, and J.~K. Furdyna,
Phys. Rev. Lett.  {\bf 89}, 127201 (2002).

\bibitem{Note4}
This order of peaks is typical for type-II QDs~\cite{Matsuda2007:APL,Bansal2009:PRB}.

\bibitem{NoteType1}
In type-I QDs CB and VB extrema coincide spatially.

\bibitem{Schoemig2003:JS}
H. Sch\"{o}mig, G. Bacher, A. Forchel, S. Lee, M. Dobrowolska, and J.~K. Furdyna,
J. Supercond.  {\bf 16}, 379 ( 2003).

\bibitem{Bacher2005:PhysE} 
G.~Bacher, H.~Sch\"{o}mig, M.~Scheibner, A.~Forchel, A.~A. Maksimov, A.~V.
Chernenko, P.~S. Dorozhkin, V.~D. Kulakovskii, T. Kennedy, and T.~L. Reinecke.
Physica E  {\bf 26}, 37 (2005).

\bibitem{Worschech2008:SST}  
L. Worschech, T. Schmidt, A. Forchel, T. Slobodskyy, G. Schmidt, and L. W. Molenkamp,
 Semicond. Sci. Technol.  {\bf 23}, 114018 ( 2008).
 
\bibitem{Klopotowski2011:PRB}  
 \L{}. K\l{}opotowski, \L{}. Cywi\'{n}ski,  P. Wojnar, V. Voliotis, K. Fronc, T. Kazimierczuk, 
A. Golnik, M. Ravaro, R. Grousson, G. Karczewski, and T. Wojtowicz,
Phys. Rev. B {\bf 83}, 081306(R) (2011).

\bibitem{Reimann2002:RMP} 
S. M. Reimann and M. Manninen, Rev. Mod. Phys. {\bf 74}, 1283 (2002); 
C. Yannouleas and U. Landman, Rep. Prog. Phys. {\bf 70}, 2067 (2007). 

\end{thebibliography}
\end{document}